# Spacecraft charging and ion wake formation in the near-Sun environment

R. E. Ergun and D. M. Malaspina

*The Laboratory for Atmospheric and Space Physics, University of Colorado, Boulder, Colorado 80309*

S. D. Bale, J. P. McFadden, D. E. Larson, and F. S. Mozer

*Space Science Laboratory, University of California, Berkeley, California 94720*

N. Meyer-Vernet and M. Maksimovic

*LESIA, Observatoire de Paris, CNRS, UPMC, Université Paris Diderot, 5 Place Jules Janssen, 92195 Meudon, France*

P. J. Kellogg and J. R. Wygant

*Department of Physics and Astronomy, University of Minnesota, Minneapolis, Minnesota, 55455*

A three-dimensional (3-D), self-consistent code is employed to solve for the static potential structure surrounding a spacecraft in a high photoelectron environment. The numerical solutions show that, under certain conditions, a spacecraft can take on a negative potential in spite of strong photoelectron currents. The negative potential is due to an electrostatic barrier near the surface of the spacecraft that can reflect a large fraction of the photoelectron flux back to the spacecraft. This electrostatic barrier forms if (1) the photoelectron density at the surface of the spacecraft greatly exceeds the ambient plasma density, (2) the spacecraft size is significantly larger than local Debye length of the photoelectrons, and (3) the thermal electron energy is much larger than the characteristic energy of the escaping photoelectrons. All of these conditions are present near the Sun. The numerical solutions also show that the spacecraft's negative potential can be amplified by an ion wake. The negative potential of the ion wake prevents secondary electrons from escaping the part of spacecraft in contact with the wake. These findings may be important for future spacecraft missions that go nearer to the Sun, such as Solar Orbiter and Solar Probe Plus.

I. Introduction

Spacecraft charging and wake formation have been a concern since the launch of the first orbiters. This concern arrises for a variety of reasons including the safety of astronauts and the safeguarding of spacecraft electrical systems[1,2]. An understanding of and, in many cases, the control of spacecraft charging also can be essential for in situ measurements of charged particles and electric fields[3-7]. For these reasons, the charging of a body immersed in a plasma has been studied over the past several decades[1-15]. We investigate spacecraft (SC) charging and wake formation under strong photoelectron fluxes. The primary motivation of this study is that NASA[16] and the European Space Agency[17] and are actively pursuing in-situ plasma measurements near the Sun.

The nominal solar wind conditions at 1 AU are such that, if a SC were at zero potential ($\Phi_{SC} = 0$), the sum of photoelectron current ($I_{ph}$), ion current ($I_I$), and secondary electron current ($I_{se}$) would exceed the thermal electron current ($I_{the}$) to a SC[1,4]. Thus, a SC typically settles to a small positive potential (a few volts), reducing photoelectron and secondary electron currents so that the net current to the SC is zero.

With a small positive potential, the solar wind electron fluxes can be measured at all energies, albeit SC potential corrections are needed for deriving the distribution function, density and velocity[6-9,18]. Since the solar wind speed is almost always supersonic, the ion fluxes have high velocities with respect to the SC, so the most meaningful part of the ion distribution can be measured as well[6-9,18]. As it turns out, the positive potential on the SC is fortunate since a negative SC could hamper the measurement of core electrons. Electric field measurements are somewhat more difficult in the solar wind since a negative potential well can form in the ion wake[3,4,8-10]. Care must be taken in design of these instruments and subsequent data analysis.

Nearer to the Sun than 1 AU, the plasma environment changes considerably. However, the most significant currents between the ambient plasma and a SC scale roughly as $\sim 1/R^2$, where $R$ is the distance from the Sun. Therefore, a SC nearer to the Sun would be expected have, under most conditions, a positive potential as well. But there are some differences between 1 AU and the near-Sun environment that do not scale as $1/R^2$. The plasma Debye length scales roughly as $R$ and electron and ion temperatures increase nearer to the Sun, so the potential in an ion wake is larger, whereas the characteristic energy of the photoelectrons and secondary electrons remain the same. These "higher-order" effects may cause the SC potential to deviate from the relatively benign potential of a few volts positive.

Analytical and numerical solutions of the charging and the plasma environment of the HELIOS SC[8,9] highlighted some of the differences in SC charging nearer to the Sun. The numerical solutions show a deeper wake potential and significant electrostatic barriers produced by the SC photoelectrons. The studies were verified by examining the electron fluxes measured on the HELIOS SC.

We investigate SC charging and wake formation in the near-Sun environment with a three-dimensional (3-D), self-consistent code. The code combines a Poisson solver and a particle tracing routine. We find that the ion wake has a negative potential that is a significant fraction of the thermal electron temperature ($kT_e$). This finding verifies previous studies[3,8-10] and punctuates that the wake effect near to the Sun is more significant than at 1 AU. The ion and electron temperatures are higher and the wake size is many Debye lengths. Another important finding is that the SC settles to a negative potential sometimes in excess of $-kT_e$, as low as ~-100 V, in spite of the high photoelectron currents. (Since there are uncertainties in secondary electron production, the floating potential of a SC cannot be predicted with high accuracy.) These findings may impact both electric field[3-5] and electron[6-9] measurements on future missions[16,17].

Further analysis shows that the negative SC potential is primarily due to an electrostatic barrier[1,8-9,11,19] which forms on the sun-exposed surfaces of the SC. The electrostatic barrier comes from a combination of conditions which include a small Debye length of the photoelec-trons ($\lambda_{ph}$) and the high thermal electron temperatures. Essentially, the thermal electrons can penetrate the electrostatic barrier whereas photoelectrons and secondary electrons cannot. Furthermore, the negative potential of the ion wake prevents photoelectron fluxes and secondary electron fluxes from escaping the surfaces of the SC that contact the ion wake[19-21].

**II. SC Charging Overview**

For a conductive SC, charging is generally solved through balancing of the currents to and from the spacecraft[1,22]:

$$I_{ph}(\Phi) + I_I(\Phi) + I_{se}(\Phi) + I_{the}(\Phi) + I_{other} = 0 \tag{1}$$

where $I_{other}$ acknowledges that there may be currents not considered here, for example, thermionic currents. For this work, we set $I_{other} = 0$. Each of these currents varies with the SC potential ($\Phi_{SC}$). The roots of the above equation yield $\Phi_{SC}$. Generally, there is only one root to this equation but multiple roots are possible if the electron distribution is non-Maxwellian and the electron secondary yield is high[1,23], or if non-monotonic potentials surround the SC[24]. If the SC is non-conducting or has isolated surfaces, each surface must be solved separately.

$I_{ph}$ depends on the projected area of the SC that is exposed to sunlight convolved with the photoelectron yield ($J_{ph0}$) which, in turn, depends on the intensity of the sunlight and the properties of the material. For most SC conducting materials, $J_{ph0}$ ranges from ~20 µA/m² to ~60 µA/m² at 1 AU[1,4,5]. Often, the photoelectron yield increases after long exposure to space vacuum over that measured in the laboratory[4,13]. We will use 20 µA/m² as a low yield and 57 µA/m² as a high yield[4] at 1 AU. The photoelectron spectrum yields a current that varies with the spacecraft potential. This relation has been previously described as a double exponential[4]:

$$J_{ph} = J_{ph0}[(1-\alpha)e^{-\Phi_{SC}/V_1} + \alpha e^{-\Phi_{SC}/V_2}], \Phi_{SC} \geq 0 \tag{2}$$

where $V_1 = 2.7$ V, $V_2 = 10$ V, and $\alpha = 5\%$. $J_{ph} = J_{ph0}$ if $\Phi_{SC} < 0$.

In the solar wind environment, $I_I$ is insensitive to the spacecraft potential in all but the most extreme cases. The solar wind velocity, ~300 km/s, is such that ions can penetrate a barrier as high as ~1 kV. $I_I$ is determined by the product of the projected area of solar wind impact, the solar wind speed, and the solar wind density. Since little is known on the absorption efficiency, we assume it to be 100%. We show later that $\Phi_{SC}$ is only moderately sensitive to the ion absorption efficiency.

To lowest order, thermal electron current is related to the thermal flux of electrons impinging on a SC:

$$J_{the0} = en\sqrt{kT_e/(2\pi m_e)} \qquad (3)$$

where $e$ is the fundamental charge, $n$ is the plasma density, and $m_e$ is the electron mass. The thermal electron current is incident to the entire exposed area of the SC. If the SC is positively charged, the electron current can increase due to focusing[22]:

$$J_{the} \cong J_{the0}(1 + \Phi_{SC}/T_e),\ 0 \leq \Phi_{SC} \ll T_e \qquad (4)$$

Equation (4) is an approximation of a sphere with a radius smaller than the Debye length. Otherwise the thermal electron current decreases:

$$J_{the} \cong J_{th0e}e^{\Phi_{SC}/T_e},\ \Phi_{SC} < 0 \qquad (5)$$

For most materials, the absorption efficiency is nearly 100% for low-energy (<50 eV) electrons[1], but can decrease above few hundred eV.

$I_{se}$, while important[20-21], is difficult to predict. The emission of secondary electrons comes from both ion impact and electron impact. The efficiencies depend on the energy of the impacting particle and are not well established for many materials. In the solar wind, the ion impact efficien-cies are expected to be close to 100%, in effect making the secondary electron current from ion impact nearly equal to $I_I$. To derive the contribution of $I_{se}$ from electron impact, one must convolve the electron fluxes with the secondary yield as a function of energy. The secondary yield is near zero for low-energy electrons (< 10 eV) but can be greater than unity if the electron energies are ~100 eV, so the net efficiency of the electron secondary emission can vary between near zero to greater than 100%[23]. The spectrum of secondary electrons typically has a characteristic energy, $V_{se}$ ~ 2 eV[1]. Thus the electron secondary currents vary with $\Phi_{SC}$ as:

$$J_{se} = J_{se0}(\Phi)e^{-\Phi_{SC}/V_{se}}, \Phi_{SC} \geq 0 \qquad (6)$$

Analytic solutions to Equation (1) are often not possible, but simple approximations can be made. For example, in the 1 AU solar wind environment, $kT_e > V_1$ and the photoelectron current is larger than all other currents. One can set $I_{the} = I_{the0}$ and assume $I_I$ and $I_{se}$ are small:

$$\Phi_{SC} \cong V_1 \ln[-I_{ph0}/I_{the}] \qquad (7)$$

where, by convention, $I_{the}$ is negative and $I_{ph0}$ is positive (current to SC is positive). This approximation yields the few volts positive potentials in the solar wind.

A solution to Equation (1) would have the underlying assumption that the paths of the charged particles to the SC are not altered by the surrounding potential as to change the net current, except as allowed in Equations (2-6). In many plasma environments, this approximation is useful but is not necessarily accurate, particularly is the SC geometry is not spherical[25]. Closer to the Sun, however, the charge density of photoelectrons is sufficient to develop non-monotonic potentials which may significantly change the currents to the SC. We investigate this behavior with a numerical code.

## III. Description of Numerical Code

A fully 3-D and a 3-D cylindrically-symmetric, Poisson solver and electron tracing program is employed to examine the potential structure surrounding a spacecraft. The primary motivation is to determine the error sources on the electric field and electron measurements from the ion wake and non-monotonic potential structures from intense photoelectron currents. An ion wake can induce a potential well on the downstream side of an object[4]. The code is a substantially modified version an earlier code used for the Cluster spacecraft[26]. It includes the ion wake, Debye shielding, and secondary electron emission as well as photo electrons.

For the fully 3-D solutions, a model of a spacecraft is placed in the center of a 20 m x 20 m x 20 m box on a 200x200x200 cubic grid with 10 cm spacing. Another version of the code has 3-D, cylindrically-symmetric domain on a 2-D grid. It allows for finer grid spacing and has significantly faster convergence. The domain is a 5 m (in $r$) x 10 m (in $x$) cylinder with 250x500 2-D grid. The grid spacing is 2 cm in both $x$ and $r$. Particle tracing is in 3-D. The codes have two parts which (1) determine the potential structure ($\Phi$) surrounding the spacecraft via a Poisson solver (given a charge distribution) and (2) determine the charge distribution via particle tracing (given $\Phi$). The two parts of the code are iterated until they converge to a self-consistent solution. Figure 1 shows the basic algorithm.

We assume that the spacecraft is conducting. We can allow for some non-conduction areas by setting them at a fixed potential, for example, the front side of the solar arrays in the Solar Probe Plus SC are assumed to be non conducting[16]. Since the fully 3-D spacecraft is constructed of 10 cm cubes, fine detail cannot be included. A thin appendage has a minimum dimension of 10 cm.

The ion density is determined by streaming ~$10^8$ ions through the box, deriving the density from the integrated dwell time inside of the grid cubes. The dominant ion motion is from the solar wind velocity (~300 km/s) and the spacecraft ram. For example, the Solar Probe Plus SC could have a velocity up to 180 km/s perpendicular to the solar wind near perihelion. The ions are initiated with the solar wind and ram speed, plus a random velocity that emulates a temperature ($T_i$), which ranges from a few eV at 1 AU to nearly 100 eV at 10 $R_S$ (solar radii) from the Sun. Ions that strike the spacecraft are removed, creating a wake in the anti-ram side of the spacecraft. This derived ion density is held fixed when deriving the solutions. One can verify *a posteriori* that the solution (potential) does not greatly change the ion current to the SC nor the ion density surrounding the SC.

The baseline thermal electron density ($n_{the0}$) is derived in a similar fashion, except that the electrons impinge on the spacecraft from all sides, creating a modest electron well surrounding the spacecraft. The resulting electron density is smoothed to remove noise. Once $\Phi$ is determined, the thermal electron density is treated as a Boltzmann fluid:

$$n_{the} = n_{the0} e^{\Phi/T_e} \qquad (8)$$

This treatment allows for self-consistent Debye shielding and is valid as long a $\Phi$ is negative or, if positive, $\Phi \ll T_e$.

The code then derives the photoelectron density by tracing photoelectrons emitted from the sunlit surfaces in the surrounding potential structure[26]. Photoelectrons (~$10^6$ particles in all) are randomly created on the sunlit surfaces with isotropic directions and an energy profile of a double exponential with characteristic energies of 2.7 eV and 10 eV (Equation 2) using a root emission of 57 µA/m² at 1 AU, scaled to the location of the SC. In some cases, a root emission of 20 µA/m² is also used for comparison with the higher emission results. Individual particles are traced by a "leap-frog" scheme in which the position is advanced then the velocity is advanced. The density is determined from the accumulated dwell time of particles within each of the grid cubes. The tracing continues until the particles either strike the spacecraft or exit the code's boundaries. Those that exit the code's boundaries are considered lost and counted in the overall photoelectron cur-rent from the plasma to the spacecraft. The secondary electron density is derived in a similar fashion, randomly creating secondary electrons over the spacecraft surface with 2 eV characteristic energy and an overall production efficiency $\varepsilon_{sec}$ estimated by convolving the electron flux energy profile with a published efficiency profile for BeCu[1, 23]. Again, we emphasize that the secondary electron production has significant uncertainty.

The spacecraft potential is estimated by balancing the currents. Once the ion, thermal electron, photoelectron, and secondary electron densities are established, the potential is derived from a Poisson solver over the entire grid, holding the spacecraft surfaces at constant potential and the boundaries at zero. The process, photoelectron tracing, secondary electron tracing, thermal electron density derivation, spacecraft potential calculation, followed by Poisson solver, is repeated until a self-consistent solution converges (Figure 1). The primary convergence criterion is the maximum change in $\Phi$ in the code's domain ($\Delta\Phi_{max}$). Depending on how many particles are used, $\Delta\Phi_{max}$ ranges from 1 mV to 25 mV.

The Poisson solvers, both the fully 3-D and 3-D cylindrically-symmetric versions, have a iterative convergence of:

$$\nabla^2\Phi = -\rho/\varepsilon_o = -(\rho_{ion} - \rho_{the} - \rho_{phe} - \rho_{se})/\varepsilon_o \qquad (9)$$

where $\rho_{ion}$, $\rho_{the}$, $\rho_{phe}$, and $\rho_{se}$, are, respectively, the ion charge density, the thermal electron charge density, the photoelectron charge density, and the secondary electron charge density. In each iteration, the value of $\Phi$ is set to $\Phi_{ave} + \rho\delta x^2/6$, where $\Phi_{ave}$ is the average of $\Phi$ in the surrounding grids and $\delta x$ is the grid spacing.

**IV. Results**

Figure 2a displays a 3-D rendering of the ion density and resulting wake ($x$ is toward the Sun, $z$ is normal to the ecliptic plane, and $y$ completes the triad) from a model of the Solar Probe Plus spacecraft at 9.5 RS from the Sun. The solar wind speed is 300 km/s (in the -x direction and the spacecraft speed is 180 km/s in the -y direction. Figure 2b shows the electron density (all electrons), Figure 2c displays the self-consistent potential field surrounding the spacecraft, and Figure 2d displays the photoelectron and secondary electron densities as a fraction of the background density ($n_0$). The plots show cuts through three planes. The expected plasma conditions at 9.5 $R_S$ were $n_0 = 7000$ cm$^{-3}$, $T_e = 85$ eV, and $T_i = 82$ eV. The high-yield photoelectron current, scaled to 9.5 $R_S$, is used, 29 mA/m$^2$. The average electron absorption efficiency is set at 85% (estimated from $T_e$), the ion absorption efficiency is set at 100%, and the electron secondary production efficiency is estimated at 100% for both ion and electron impacts.

The potential in the center of the wake is below -60 V (Figure 2c) due to the ion vacuum (Figure 2a). The deep ion wake potential is expected since the scale size of the wake, ~ 2-3 m in diameter, is larger than the thermal electron Debye length ($\lambda_D = 0.76$ m). The center of the wake is expected to see potentials on the order of $T_e$. The temperature of the thermal electrons is such that they easily penetrate the electrostatic barrier surrounding the spacecraft, but are significantly altered in the ion wake region (Figure 2b). They also have a significant shielding effect. The plasma potential ~5 m from the SC is nearly zero.

Figure 2c also shows a thin layer of negative potential surrounding the SC. It is particularly strong on the sunward side of the spacecraft due to the high photoelectron density (>10$^6$ cm$^{-3}$), also in a thin layer (Figure 2b and Figure2d). The non-monotonic, electrostatic barrier prevents the low-energy part of photoelectrons and secondary electrons from escaping. On the top of the SC, the electrostatic barrier is ~ -8 V and reflects ~99% of the photoelectron current and >99% of the secondary electrons back to the SC. A smaller (< 1 V) barrier is seen around the other SC surfaces. These barriers can cause the SC to have negative potential.

Figure 2d also shows that secondary electrons do not escape from the areas of the SC that are contacting the ion wake. The secondary production is smaller since less thermal electrons reach the surface contacting the wake and the negative wake potential (-60 V) very efficiently reflects these secondary electrons (~2 eV) back to the SC. This wake reflection amplifies the negative potential of the SC.

In the 3-D solutions, the electrostatic barrier is mostly carried in one grid layer, so it is examined further with a series of 3-D, cylindrically-symmetric numerical solutions of simple, fully-conducting cylinder, 1 m in radius and 2 m long, with one end allowed to emit photoelectrons. This controlled experiment allows us to investigate the conditions that cause the negative charging.

Figure 3 shows a solution that has no ion wake. The ions are artificially held at a fixed density. This condition is unrealistic, but is useful to examine the effect of the ion wake. The plasma conditions are otherwise identical to those in Figure 2 as are the electron absorption and electron secondary emission efficiencies. Figure 3a displays $\Phi$ and Figure 3b displays the $n_{ph} + n_{se}$. $\Phi_{SC}$ is -0.85 V. The currents to the SC are in Table 1. In Figure 3a, one can clearly see an electrostatic barrier on the top face of the cylinder, the face that has photoelectrons. A smaller electrostatic barrier also surrounds the SC on all sides. This negative barrier comes from the large $n_{ph}$ (1.1 x $10^6$ cm$^{-3}$) and $n_{se}$ (2.9 x $10^4$ cm$^{-3}$) that forms in a thin layer around the SC.

Figure 4a plots the potential along the $r = 0$ axis of the solution. The electrostatic potential on the top of the SC ($x = 1.15$ m) is -6.9 V causing a barrier of approximately -6.1 V with respect to $\Phi_{SC}$. Effectively, this electrostatic barrier reflects 92% of the photoelectrons back to the SC (Table 1). Figure 4b plots $n_{ph}$ (red), $n_{se}$ (purple), $n_{the}$ (blue), and their sum $n_e = n_{ph} + n_{se} + n_{the}$ (black) along the same axis. The vertical dashed lines are the edges of the SC. The thin layer of high electron charge is seen at both ends of the SC. The charge layer at the $x = 1$ m end, is primarily from SC photoelectrons. Using $n_{ph} = 10^6$ cm$^{-3}$ and $T_{ph} = 3$ eV, the photoelectron Debye length is ~15 cm, the thickness of the electrostatic barrier. The location of the barrier, about a photoelectron Debye length from the SC, is roughly that predicted from analytic solutions of a flat plate[24]. The depth of the electrostatic barrier (~- 6 V) is such that the escaping photoelectrons are from the higher-energy tail (Equation 2).

Figure 4a also shows a mild electrostatic barrier on the $x = -1$ m side of the SC. The minimum potential at x = -1.15 m forms a barrier of ~-0.3 V. This barrier, due to secondary electrons, causes approximately a 15% loss of secondary electron current. Figure 4a shows an overshoot of the potential at x = -2.3 m and x = 2.5 m. This overshoot is due to the shadowing of the thermal electrons by the spacecraft whereas the ion density is fixed at $n_0$. In the solar wind environment, a fixed ion density is realistic on the +x side of the SC, but one would expect an ion wake on the -x side of the SC.

Figure 5a shows the ion density with an ion wake on the -x side of the SC. The ion temperature is ignored, so the wake has a complete ion vacuum. Figure 5b displays the thermal electron density, Figure 5c displays $\Phi$, and Figure 5d displays $n_{ph} + n_{se}$. The SC potential is - 4.15 V. $\Phi$ reaches -37 V in its center of the ion wake. The secondary electrons cannot escape from the -x side of the SC (Figure 5d), so the secondary electron current is decreased (Table 2). Comparing the solutions in Figure 3 and Figure 5, one can see that the ion wake amplifies the negative charging of the SC. In this example, the areas of the SC from which secondary electrons cannot escape are the top and bottom ends, which are about 1/3 of the total area of the SC. The ion wake and the electrostatic barrier from the photoelectrons in the Solar Probe example (Figure 2) blocks over one half of the SC area, causing the $\Phi_{SC}$ to have a more negative value.

Figure 6 displays the results of a solutions with conditions four times farther from the Sun, $n_0$ = 440 cm$^{-3}$, $T_e$ = 25 eV, and the photoelectron yield reduced by a factor of sixteen, to 1.8 mA/m$^2$. The same absorption and production efficiencies are used. Figure 6a displays $\Phi$ and Figure 6b shows $n_{ph} + n_{se}$. In this case, $\Phi_{SC}$ = 2.9 V. The electrostatic barrier at the top of the SC is much weaker. In particular, photoelectrons have a clear path radially outward. If $\Phi_{SC} = 0$, $I_{ph}$ would greatly exceed the $I_{the}$, so the SC charges to a positive potential as often seen at 1 AU.

As a further test, a solution is found for a 1/4 scale model of the SC under the same plasma conditions as in Figure 5. Figure 7 displays the results. The spacecraft is 0.25 m in radius and 0.5 m long and the code's domain is also 1/4 in size, 1.25 m in $r$ and 2.5 m in $x$. The SC charges to a small, positive potential (0.3 V).

**V. Discussion and Conclusions**

The 1/4-scale model of SC floats to a positive potential in the same plasma conditions that the full-scale model has a negative potential. This conspicuous difference can be understood by comparing the scale size of the electrostatic barrier ($\lambda_{Dph}$) with the size of the SC ($R_{SC}$). If $R_{SC} \gg \lambda_{Dph}$, the photoelectron current essentially can be treated as a 1-D problem[24]. Near the center ($r = 0$) of the top of the SC ($x = 1$ m), $n_{ph} \gg n_{the}$, $n_I$, and $n_{se}$, causing a high negative charge layer, so $\Phi$ falls rapidly with distance (along $x$) from the SC. Therefore, $n_{ph}$ falls rapidly

$$n_{ph}(x) = n_{ph0} e^{(\Phi(x) - \Phi_{SC})/T_{ph}} \qquad (10)$$

where $T_{ph}$ is a characteristic temperature of the photoelectrons. This decrease in $\Phi$ and $n_{ph}$ can occur even if $\Phi_{SC}$ is negative. This effect is seen in Figure 4.

$\Phi(x)$ continues to decrease with $x$ until, $n_{ph} = n_I - n_{the} < n_I$, assuming $n_{se}$ is negligible. Near the SC, $n_{the}$ is less than $n_I$ because the electron fluxes are partially physically screened by the SC whereas the ions fluxes are not screened except in the ion wake. In a 1-D solution, the electrostatic barrier builds until $n_{ph} = \alpha n_I$, where $\alpha < 1$. Charge balance will occur at a distance from the SC of several $\lambda_{Dph}$. If $R_{SC} \leq \lambda_{Dph}$, the problem is treated in 2-D or 3-D, so $n_{ph}$ will naturally fall with distance from the SC with or without the electrostatic barrier. For this reason, SC at 1 AU show a much milder barrier[11].

Ultimately, if $n_{ph}$ is limited to a fraction of $n_I$, the limiting photoelectron current from the SC is $J_{phL} = \alpha e n_I v_{ph}$, where $v_{ph}$ is a characteristic velocity of the photoelectron flux at the limiting point. With a significant barrier, $v_{ph}$ will be ~$10^6$ m/s, the speed of a 10 eV electron due to the high-energy tail in the photoelectron fluxes. Since the thermal electron speed is larger, $J_{phL} \ll J_{the} = e n_I v_{the}$. We conclude that if (1) $R_{SC} \gg \lambda_{Dph}$ and (2) $v_{the} > \alpha v_{ph}$, the thermal electron current will exceed the photoelectron current so the SC may charge to a negative potential. The solution in Figure 5 has $R_{SC} = 1$ m and $\lambda_{Dph} \sim 0.15$ m, whereas the solution in Figures 6 and 7 have $R_{SC}$ and $\lambda_{Dph}$ nearly equal. The model of the Solar Probe Plus SC has $R_{SC} \gg \lambda_{Dph}$.

Interestingly, if $R_{SC} \gg \lambda_{Dph}$, the escaping photoelectron current does not strongly depend on the photoelectron yield and the surface of the SC ($J_{ph0}$), as long as it is sufficient to form an electrostatic barrier. The solution in Figure 5 was only slightly different using the low-yield value of $J_{ph0}$.

The SC will not necessarily charge to a negative potential if $I_{se}$ is comparable to $I_{the}$. However, the numerical solutions show that, even under high secondary yields (100%), the SC can charge to a negative potential because $I_{se}$ is reduced by (1) the electrostatic barrier on the top of the SC caused be the photoelectrons, (2) a small barrier caused by the secondary electrons, and (3) the negative potential of the ion wake. These reductions cause the model of the Solar Probe Plus SC to float to ~ -10 V in our solutions. With lower secondary yield (50%), the negative charging becomes more severe and can be lower than $\Phi_{SC} \sim -kT_e$ (-85 V).

We examined the possibility of multiple roots[1,23-24] in Equation (1). Multiple roots are possible if the electron distribution is non-Maxwellian and the electron secondary yield is high[1,23], or if non-monotonic potentials surround the SC[24]. We used the same Poisson-based code to search for multiple roots from non-monotonic potentials by fixing $\Phi_{SC}$, forcing a solution, then recording the net current to the SC. (The distributions are modeled as Maxwellian, so multiple roots due to non-Maxwellian distributions could not be uncovered.) We could not find evidence for multiple roots. We, however, cannot rule out multiple roots since we examined only a small number of cases. The possibility of multiple roots is a subject for future research.

In conclusion, numerical solutions show that the SC can charge to a negative potential in spite of very high photoelectron fluxes. This behavior can be understood by comparing the size of the SC with the photoelectron Debye length. If $R_{SC} \gg \lambda_{Dph}$ and $T_e$ is significantly larger than the characteristic energy of photoelectrons, an electrostatic barrier can form on the sunlit surfaces that reflects part of the photoelectron flux back to the SC. The negative SC potential is amplified by the fact that the secondary electron fluxes cannot penetrate the same electrostatic barrier and, if an ion wake forms, they cannot escape from the area of the SC that contacts the ion wake. Depending on the secondary electron yield, the $\Phi_{SC}$ can range from a few Volts negative to as much as $\Phi_{SC} \sim -kT_e$ on a model of the Solar Probe Plus SC. This charging could compromise the measurement of the electron distribution and electric fields.

**Tables**

**Table 1: Currents To SC**

| No Wake (Figure 3,4) | ($\Phi_{SC}=0$) Current (mA) | Efficiency | Predicted Current ($\Phi_{SC}=0$) | Numerical Solution ($\Phi_{SC}=-0.85$) |
|---|---|---|---|---|
| $I_{ph}$ | 91.1 | 29 mA/m² | 91.1 | 7.4 |
| $I_{the}$ | -18.0 | 85% | -15.3 | -15.2 |
| $I_I$ | 1.4 | 100% | 1.4 | 1.4 |
| $I_{se}$ | 16.7 | 100% | 16.7 | 6.4 |

**Table 2: Currents To SC**

| Ion Wake (Figure 5) | ($\Phi_{SC}=0$) Current (mA) | Efficiency | Predicted Current ($\Phi_{SC}=0$) | Numerical Solution ($\Phi_{SC}=-4.15$) |
|---|---|---|---|---|
| $I_{ph}$ | 91.1 | 29 mA/m² | 91.1 | 7.8 |
| $I_{the}$ | -18.0 | 85% | -15.3 | -14.6 |
| $I_I$ | 1.4 | 100% | 1.4 | 1.4 |
| $I_{se}$ | 16.0 | 100% | 16.0 | 5.4 |

**Figures**

**Figure 1**. The basic architecture of the Possion/electron tracing code.

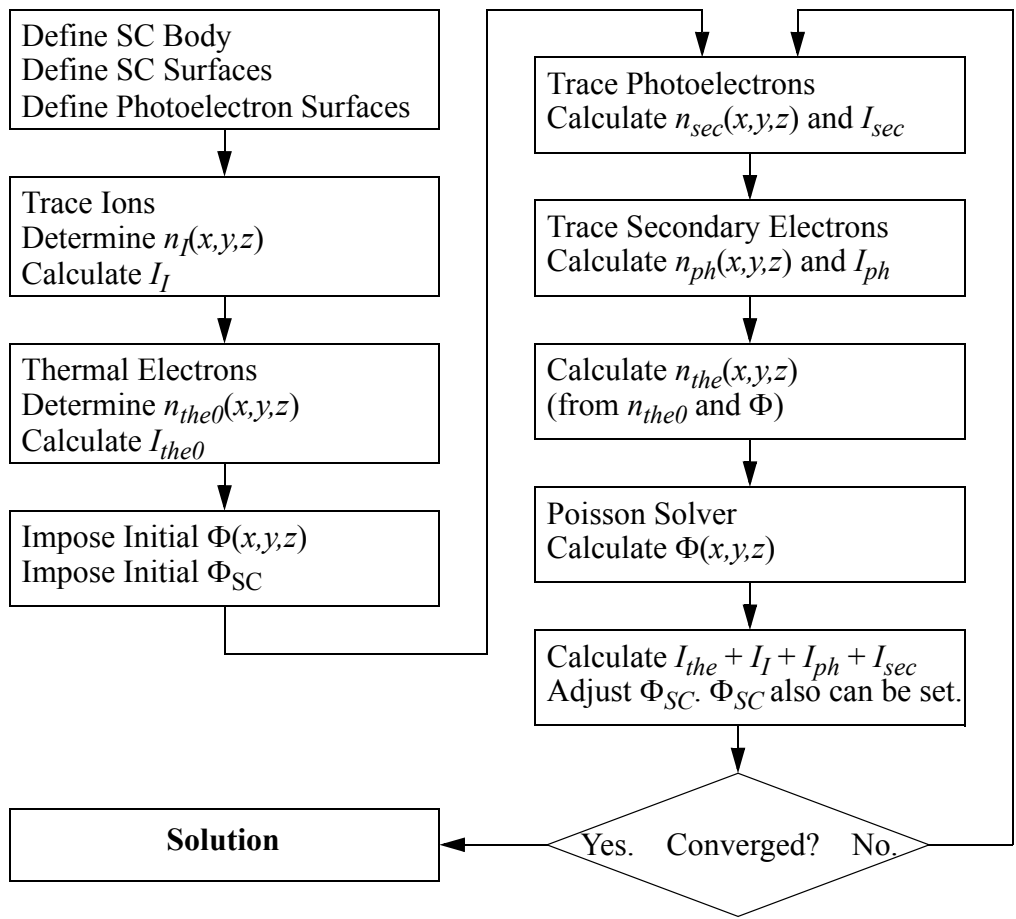

**Figure 2**. (Color) (a) A 3-D rendering of the ion density surrounding a model of the Solar Probe Plus spacecraft at 9.5 $R_S$. $x$ is toward the Sun, $z$ is normal to the ecliptic plane, and $y$ completes the triad. The distances are in meters. The the solar wind speed is 300 km/s in the $-x$ direction and SC is traveling at 180 km/s in the $-y$ direction. The plasma density is 7000 cm$^{-3}$ and the ion temperature is set at 82 eV. (b) The electron density, $n_e = n_{ph} + n_{se} + n_{the}$. The thermal electrons ($T_e = 85$ eV) dominate except for a thin layer surrounding the SC. (c) The self-consistent potential. $\Phi_{SC}$ is ~-10V. The potential well in the bottom left is created by the ion wake. A thin layer if negative potential is surrounding the spacecraft; it is particularly strong on the sunward side of the spacecraft. (d) The photoelectron and secondary electron density. All but ~1% of the photoelectrons are reflected back to the SC. The ion wake prevents secondary electrons from escaping form the left side.

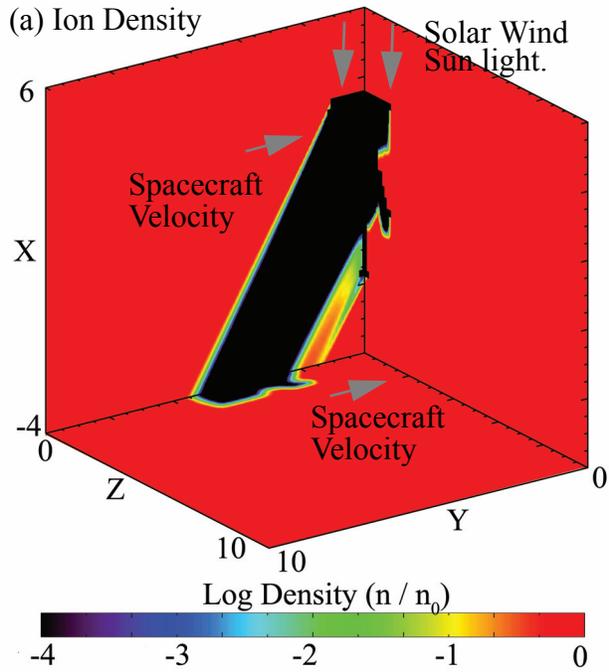
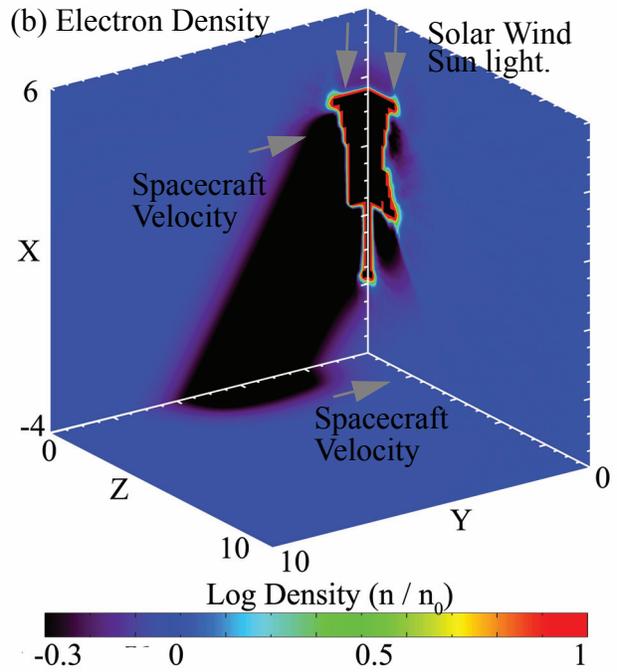
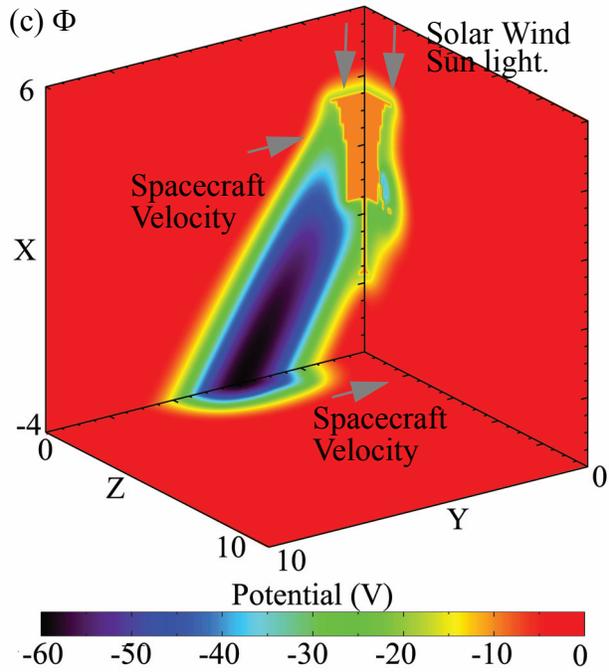
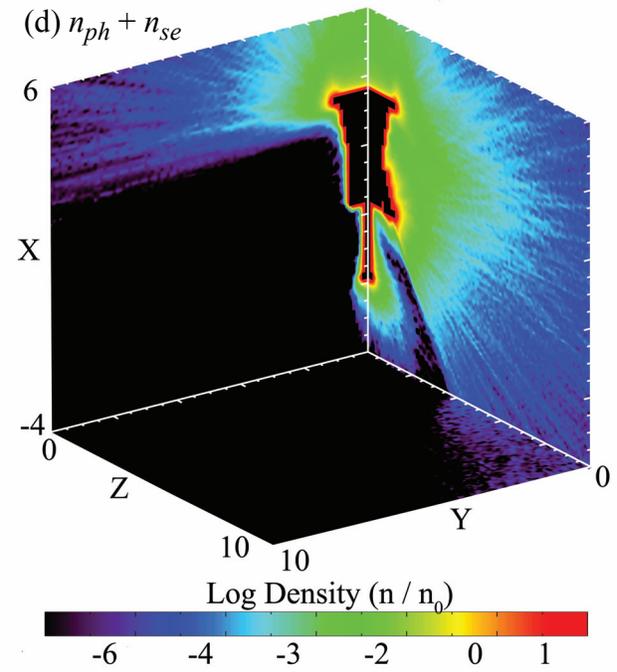

**Figure 3**. (Color) A 3-D, cylindrically-symmetric solution of a cylindrically-shaped SC at 9.5 $R_S$. The photoemission and electron distribution mimic the solar wind at 9.5 $R_S$, but the ion density is fixed. (a) Φ surrounding the SC. $Φ_{SC}$ is -0.85 V. A negative potential envelops the surface of the SC and a barrier is formed at x = +1.15 m. The ~+ 1 V structure ~ 1 m from the SC comes from a loss of thermal electron density due to partial phyiscal shielding by the SC. (b) The photoelectron and secondary electron density. A thin layer of high electron density surrounds the SC.

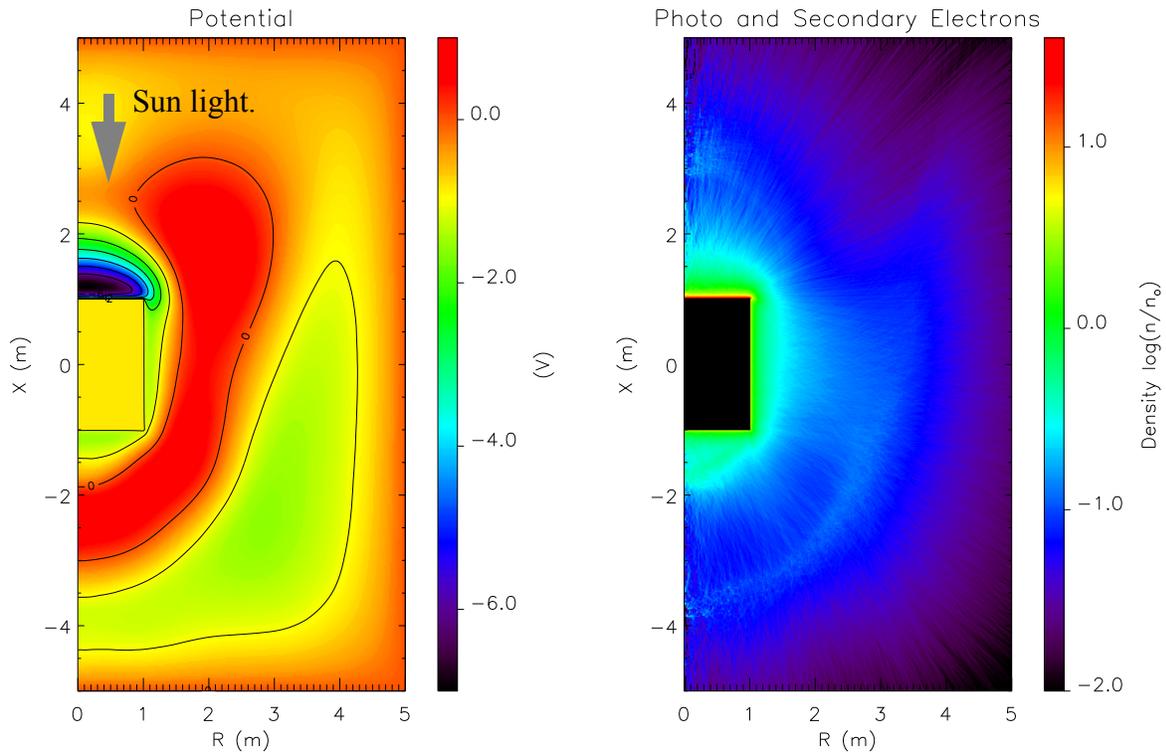

**Figure 4.** (a) A line plot of $\Phi(x)$ along the $r = 0$ axis of the solution derived from Figure 3a. The potential of the SC is -0.85 V. A -6V (with respect to $\Phi_{SC}$) electrostatic barrier forms at $x = +1.15$ m. A smaller barrier is seen at $x = -1.15$ m. (b) The electron densities along the $r = 0$ axis. The total electron density, $n_e = n_{ph} + n_{se} + n_{the}$, is in black, $n_{ph}$ is red, $n_{se}$ is purple, and $n_{the}$ is blue. The vertical dashed lines indicate the top and bottom surfaces of the SC.

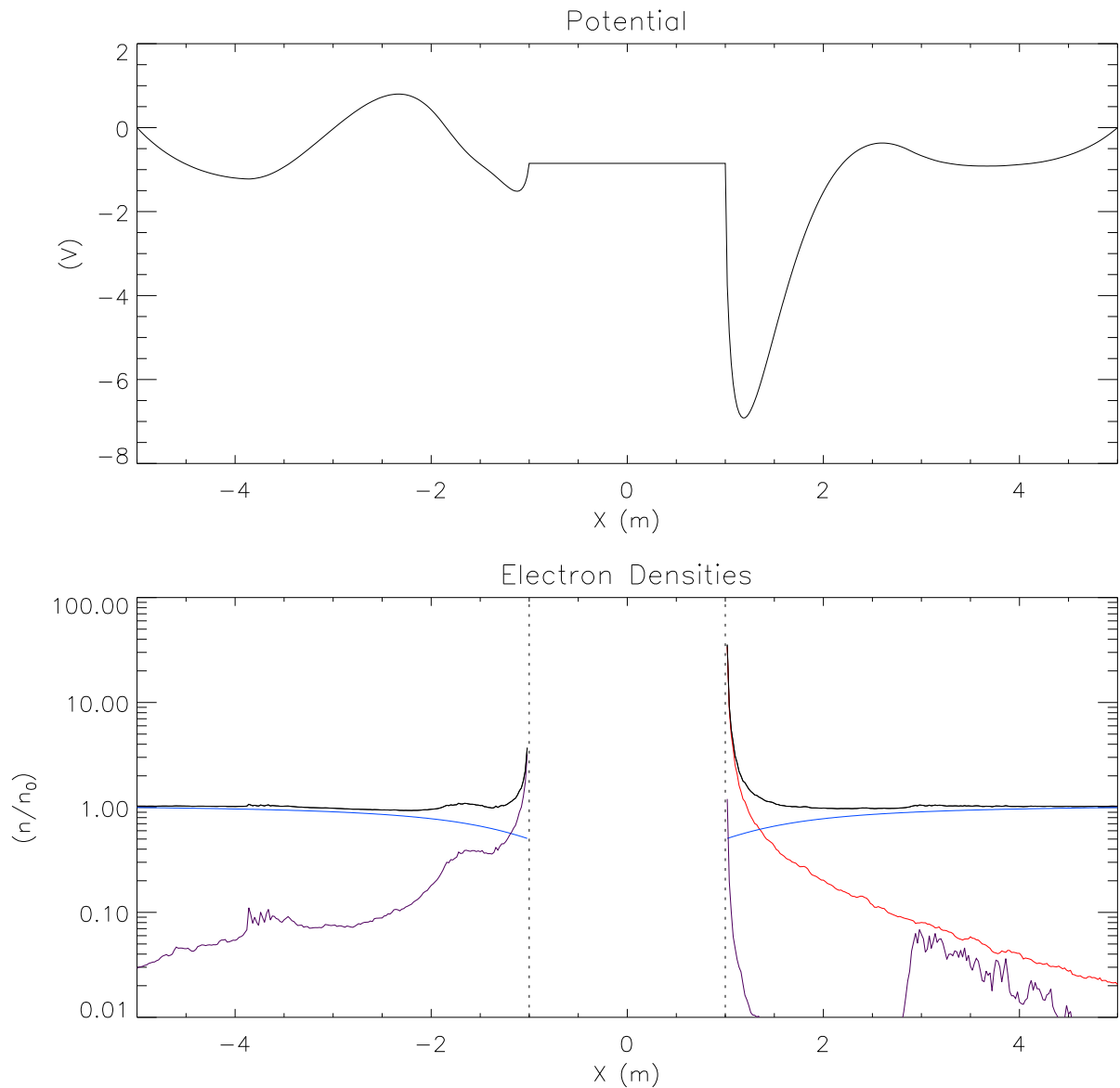

**Figure 5**. (Color) (a) The ion density. A wake is on the *-x* side of the SC. The ion temperature is zero. (b) The thermal electron density. (c) The self-consistent solution of Φ. (d) The photoelectron and secondary electron density. Secondary electrons cannot escape from the bottom of the SC.

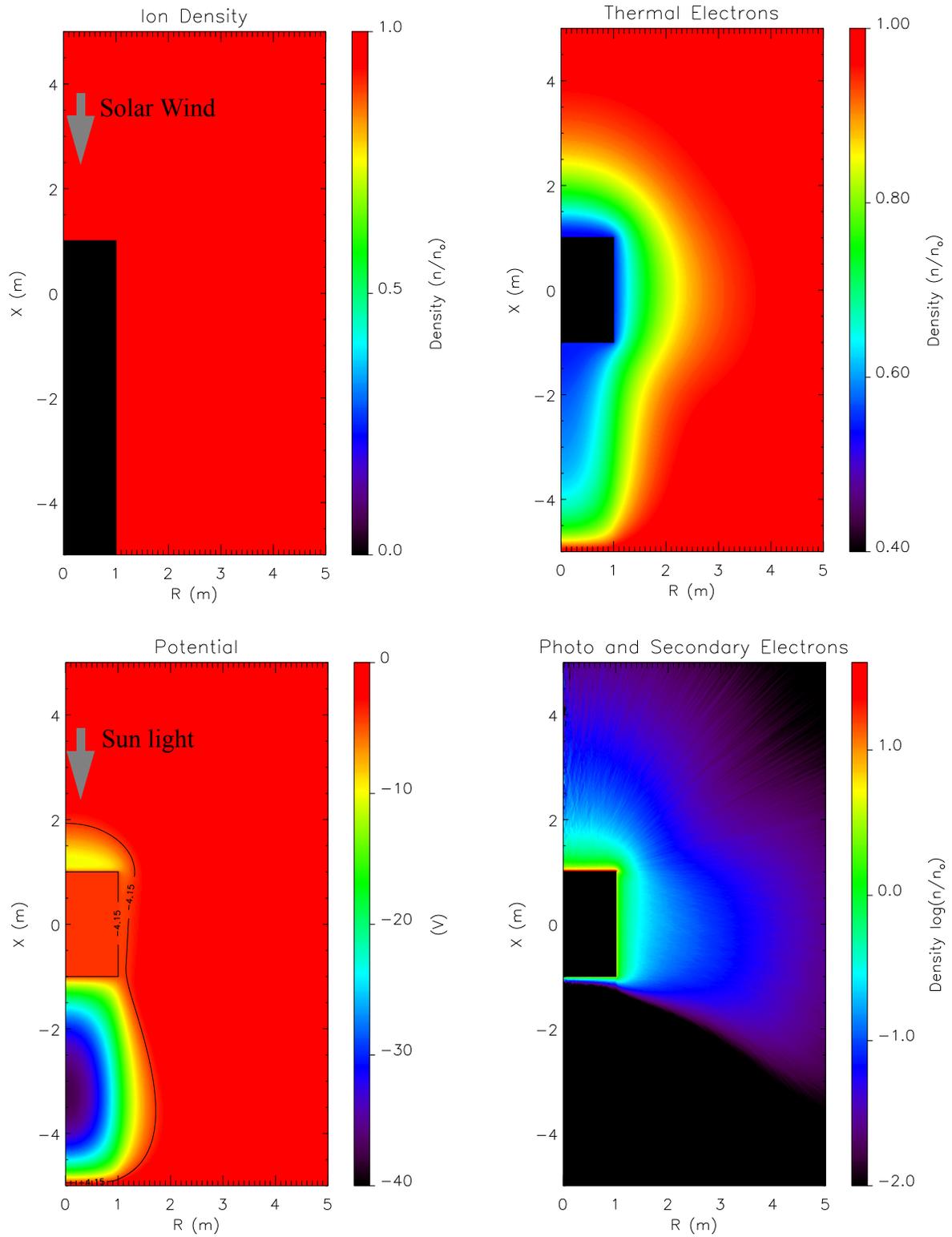

**Figure 6**. (Color) The solution under conditions four times farther from the Sun. $n_0 = 440$ cm$^{-3}$, $T_e$ = 25 eV, and the photoelectron yield is reduced by a factor of sixteen, to 1.8 mA/m$^2$. (a) Φ. The SC is at 2.9 V. (b) The photoelectron and secondary electron density.

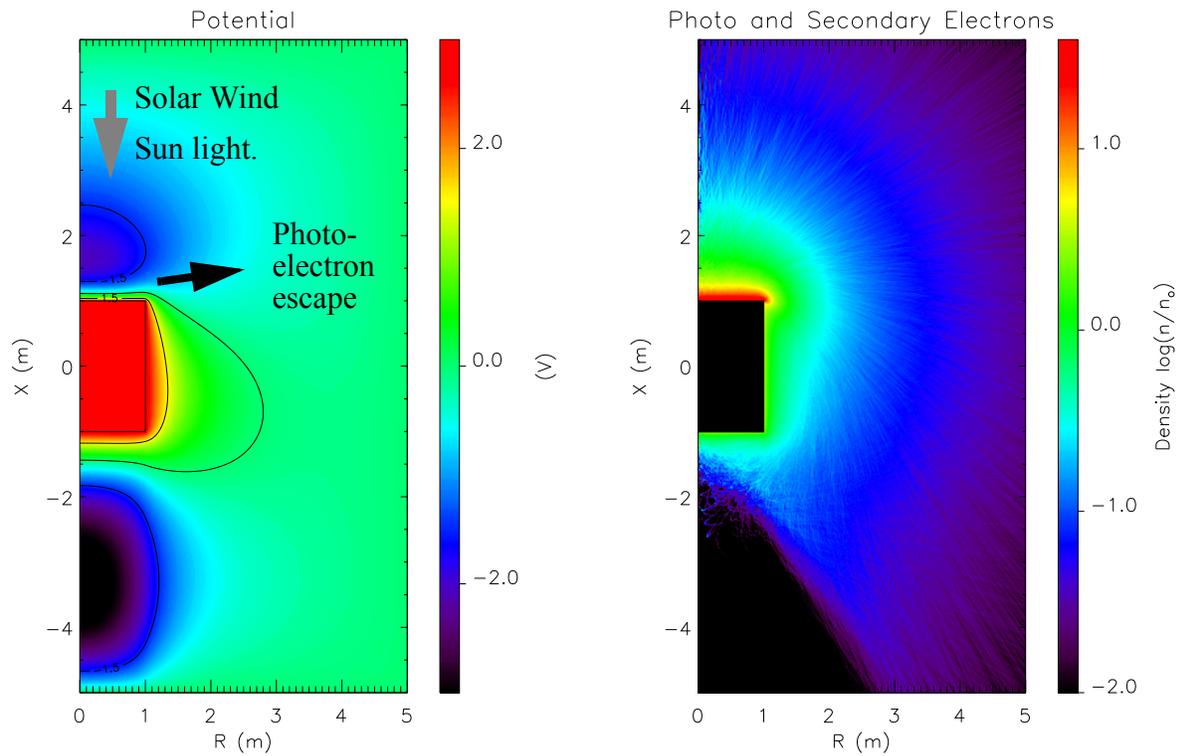

**Figure 7**. (Color) The solution of a scaled model (1/4 size) with the same conditions as in Figure 5. $n_0 = 7000$ cm$^{-3}$, $T_e = 85$ eV, and the photoelectron yield is 29 mA/m$^2$. The spacecraft is 0.25 m in radius and 0.5 m long. The solution's domain is 1/4 size as well. (a) Φ. The SC is at 0.3 V. (b) The photoelectron and secondary electron density.

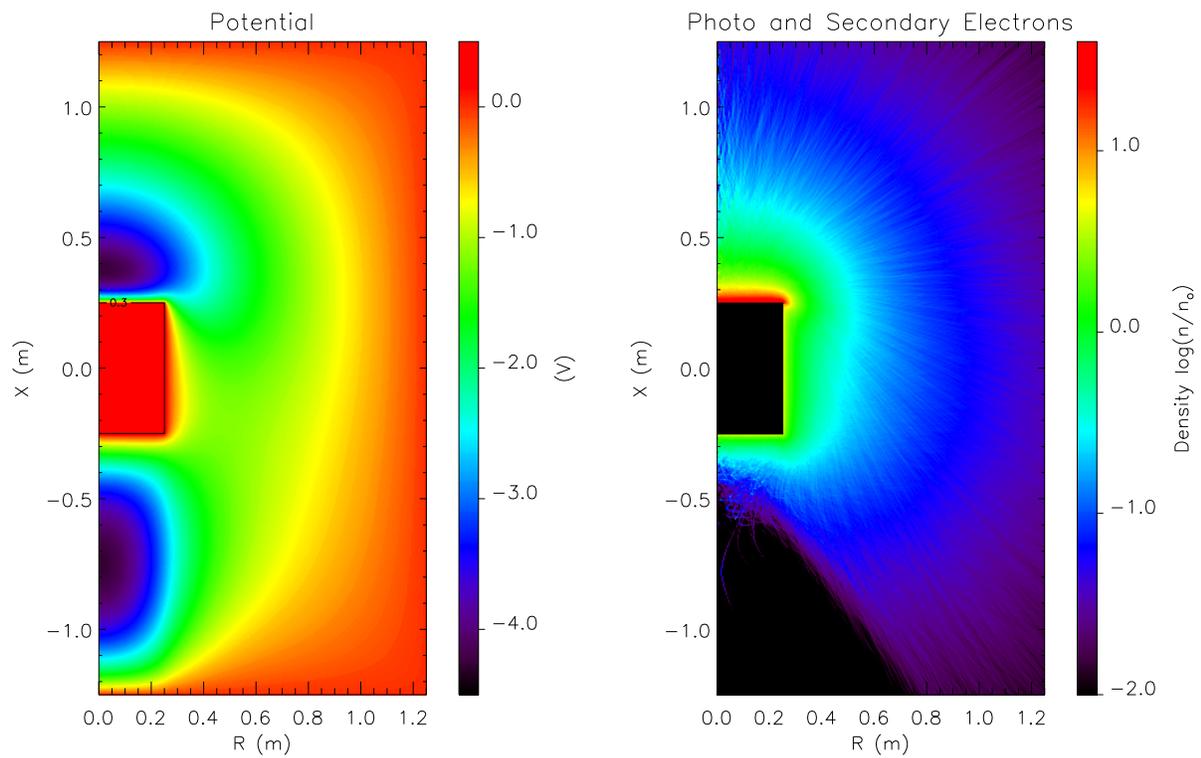